# HEALTH DETECTION ON CATTLE COMPRESSED IMAGES IN PRECISION LIVESTOCK FARMING

Sebastián Tapias Gómez
Universidad Eafit
Colombia
stapiasg1@eafit.edu.co

Valeria Cardona Velásquez
Universidad Eafit
Colombia
vcardonav@eafit.edu.co

Miguel Angel Calvache Giraldo
Universidad Eafit
Colombia
macalvachg@eafit.edu.co

Simón Marín
Universidad Eafit
Colombia
smaring1@eafit.edu.co

Mauricio Toro
Universidad Eafit
Colombia
mtorobe@eafit.edu.co

**ABSTRACT**

The constant population growth brings the needing to make up for food also grows at the same rate. The livestock provides one-third of human's protein base as meat and milk.[1] To improve cattle's health and welfare the pastoral farming employs Precision Livestock farming (PLF). This technique implementation brings a challenge to minimize energy consumption due to farmers not having enough energy or devices to transmit large volumes of information at the size are received from their farm's monitors. Therefore, in this project, we will design an algorithm to compress and decompress images reducing energy consumption with the less information lost. Initially, the related problems have been read and analyzed to learn about the techniques used in the past and to be updated with the current works. We implemented Seam Carving and LZW algorithms. The compression of all images, around 1000 takes a time of 5 hours 10 min. We got a compression rate of 1.82:1 with 13.75s average time for each file and a decompression rate of 1.64:1 and 7.5 s average time for each file. The memory consumption we obtained was between 146MB and 504 MB and time consumption was between 30,5s for 90MB to 12192s for 24410 MB, it was all files.

## 1. INTRODUCTION

In 2015 almost 1.7 million cattle were lost to nonpredator causes. Respiratory problems accounted for the highest percentage of losses (23.9 percent), followed by unknown nonpredator causes (14.0 percent) and old age (11.8 percent). [2]. Some of these deaths can be prevented with accurate and real-time information about the behavior, health, and movements of cattle. This issue is one of the multiple reasons why emerged the Precision Livestock Farming which implements information and communication technology. It optimizes in different ways the cattle rearing, offering continuous and automated information about the animals to detect health issues at an early stage or even improve cattle's welfare, livestock economy, sustainability, and others.[3]

Monitoring cattle's health through an algorithm which uses pictures of the animals to distinguish healthy from sick cattle is one of the best ways to improve efficiency related to a better farming process. Due to the number of animals in these farms, such algorithm would consume lots of energy. A way to reduce this consumption is by compressing the pictures taken to be then processed.

### 1.1. Problem

The problem we are facing is the number of images and the energy consumption it takes to transmit data in the farm to devices or cloud to get process and at that point works with precision livestock farming. Consequently, we will design an algorithm that compress and decompress images with the minimum amount of energy consumption possible. Hence, the algorithm will be evaluated through a classification animal health algorithm with the goal to reach the maximum compression rate with the classification accuracy affected below 5%.

### 1.2 Solution

In this work, we used a convolutional neural network to classify animal health, in cattle, in the context of precision livestock farming (PLF). A common problem in PLF is that networking infrastructure is very limited, thus data compression is required.

To solve the problem, we first implemented a lossy compression algorithm (Seam Carving) to reduce the size and quality of the images with a high percentage to reduce the processing cost. This algorithm was implemented because it lets to configure the number of seams to apply to compress the image. It means that the ratio compression can be adapted as we wanted. So, this algorithm was efficient to use when compressed images are required for a classification algorithm, this implies saving memory and time. The loss of information in this procedure will be the minimum possible in order not to lose effectiveness in the convolutional processing network. After this compression, we will use a lossless algorithm (LZW), which will further compress the image without losing information and save memory space and time to work with the images in the convolutional classification network.

### 1.3 Article structure

In what follows, in Section 2, we present related work to the problem. Later, in Section 3, we present the data sets and methods used in this research. In Section 4, we present the algorithm design. After, in Section 5, we present the results. Finally, in Section 6, we discuss the results and we propose some future work directions.

## 2. RELATED WORK

In what follows, we explain four related works on the domain of animal-health classification and image compression in the context of PLF.

### 2.1 A systematic literature review on the use of machine learning in precision livestock.

In this article, the authors focus on getting all information related to the recent use of machine learning in PLF and what are the techniques to collect and process data, which is fundamental on taking better decisions to drive farming into a healthier (not only for cattle but for the environment) and more productive industry. With all collected and analyzed data, the article supports some of the best ways to gain advantage from machine learning (ML) to use it in PLF [4]. Due to the nature of this work, such thing as an algorithm wasn't used to solve the problem of gathering and choosing the better uses of ML in PLF.

### 2.2 An Animal Welfare Platform for Extensive Livestock Production Systems.

Recently, the USA reformed its agricultural policy to support livestock well-being. Keeping up with new technologies and EU requirements, the authors introduce a recent development of wireless sensors, a collar device designed at a low cost. It records different parameters of animal's well-being as movement, geolocalization, and animal behavior in livestock farms. Since the battery life was a challenge, the device works in two radio frequencies, the first one can transmit real-time data in a short-range and fast, but it takes more energy consumption. The second one is slower, but its range to transmit data information is wider and, its energy consumption is lower than the first, also, works offline. The system developed can process pattern recognition through Deep Neural Network algorithms. In addition, it uses cloud computing for both data and Deep Learning model storage and usable and effective visualizations in mobile devices that provide farmers with valuable information.[5]

### 2.3 Cloud services integration for farm animal's behavior studies based on smartphones as an activity sensor.

The authors studied the iPhone sensors for tracking livestock and the challenges sensors presented. First, the data size must be reduced to minimize energy consumption. Next, the storage and processing of a large amount of data. Finally, matching multiple data from different sources.

They developed a data storage architecture able to share, treat, collect, and process from other platforms. Also, they used a UDP (User Datagram Protocol) protocol on WIFI to transmit data, but this could lead to data loss problems if a large amount of data is sending at the same time. The data compression was performed in two ways; eliminating redundancies, that is replacement of redundant data by a time interval during which the value remains constant was applied to preserve data integrity, and truncating data to 3, 4, and 5 decimal digits. [6]

### 2.4 Visual Localisation and Individual Identification of Holstein Friesian Cattle via Deep Learning

Since holstein fresian cattle are the highest milk-yielding bovine type [7] and many countries have taken rule against the traditional ear-tagging; identification and traceability look like a new challenge for farms and industry. Under this context, and aiming to a sharp technology solution, a localization using cameras and algorithms to identify these fresian bovine seems to be one of the best ways to keep tracking the cattle.

After taking the photos of the cattle from a UAV, some neural networks such as LSTM, R-CNN, Inception V3 are used to count the correct type of bovines. [8]

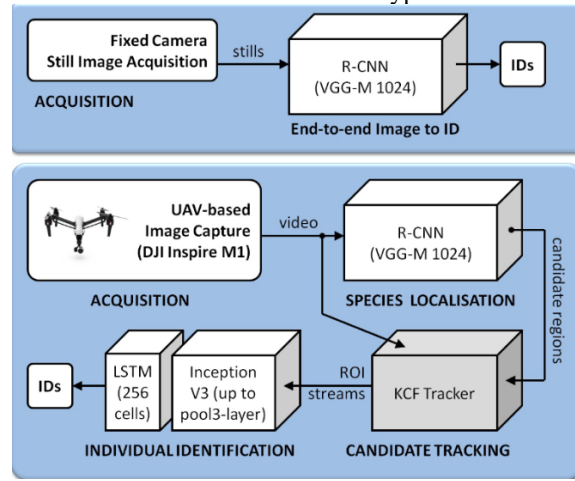

*Image 1. Visual localisation and individual identification of holstein friesian cattle via deep learning.*

## 3. MATERIALS AND METHODS

In this section, we explain how the data was collected and processed and, after, different image-compression algorithm alternatives to solve improve animal-health classification.

### 3.1 Data Collection and Processing

We collected data from Google Images and Bing Images divided into two groups: healthy cattle and sick cattle. For healthy cattle, the search string was "cow". For sick cattle, the search string was "cow + sick".

In the next step, both groups of images were transformed into grayscale using Python OpenCV and they were transformed into Comma Separated Values (CSV) files. It was found out that the datasets were balanced.

The dataset was divided into 70% for training and 30% for testing. Datasets are available at https://github.com/mauriciotoro/ST0245-Eafit/tree/master/proyecto/datasets .

Finally, using the training data set, we trained a convolutional neural network for binary image-classification using Google Teachable Machine available at https://teachablemachine.withgoogle.com/train/image.

### 3.2 Lossy Image-compression alternatives

In what follows, we present different algorithms used to compress images.

### 3.2.1 Seam Carving

Seam carving algorithm creates seams from images to later eliminate those which have less information (those which have less energy on the energy map created from the picture). Eliminating some of the less important paths on the image, it resizes and becomes lighter. [9]. The algorithm complexity is O(m,n), which m, n represent width and length of pixeled image respectively.

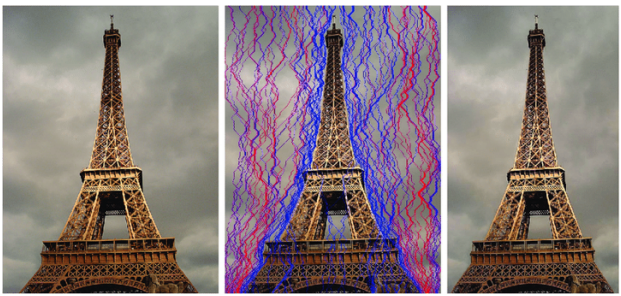

*Image 2. Seam Carving*

### 3.2.2 Discrete cosine transform or DCT

The DCT works by separating images into parts of differing frequencies. During a step called quantization, where part of compression occurs, the less important frequencies are discarded, hence the use of the term "lossy". Then, only the most important frequencies that remain are used retrieve the image in the decompression process. The algorithm complexity is $O(N\ log_2 N)$[10].

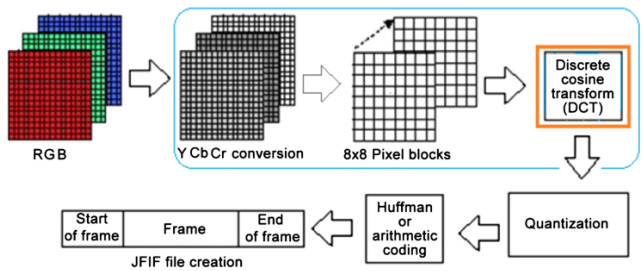

*Image 3. Discrete Cosine Transform*

### 3.2.3 Fractal compression

Fractal compression divides an image into sub-blocks. Then, it searches similarities in each sub-block. If found, replace a region with another similar one. This basically storages less information since a single block is used in two of them. The algorithm compression in the worst case is $O(N^3)$[11]

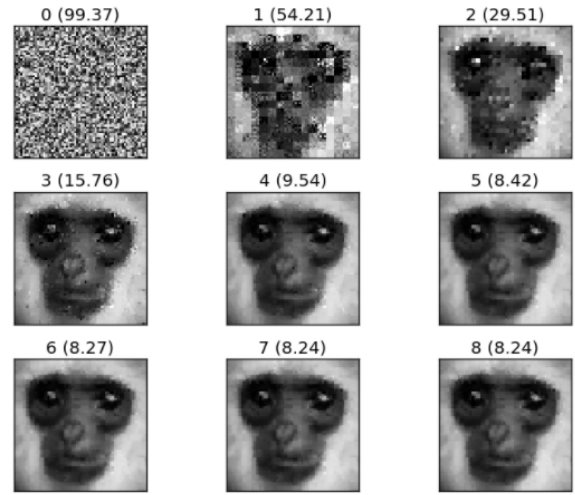

*Image 4. Fractal Compression (Decompression)*

### 3.2.4 Wavelet compression

This algorithm is a variant of DCT. First, a wavelet transform is applied to the image. This produces as many coefficients as there are pixels in the image. These coefficients can then be compressed more easily because the information is statistically concentrated in just a few coefficients. After that, the coefficients are quantized, and the quantized values are entropy encoded and/or run length encoded. The algorithm compression in the worst case is $O(N^2)$ [12]

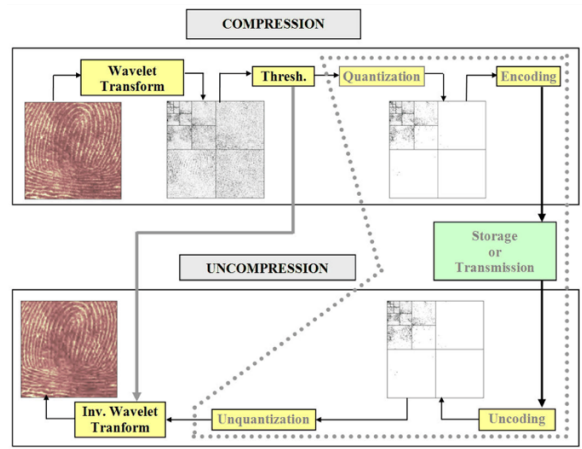

*Image 5. Wavelet compression*

### 3.3 Lossless Image-compression alternatives

In what follows, we present different algorithms used to compress images.

### 3.3.1 Burrows- Wheeler transformation

The burrows-Wheeler (BWT) permutes the positions of a string of characters to a string of repeated characters. Its data structure is the strings. The algorithm is useful because is easy to compress repeated character strings by techniques as move-to-front transform and run-length encoding. Moreover, BWT is reversible. It means the encoded string can be reconstructed to the initial word which was transformed. The algorithm complexity in the worst case is O(n) [13]

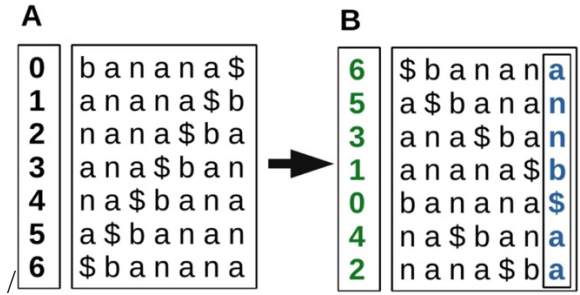

*Image 6. Burrows- Wheeler transformation example*

### 3.3.2 Huffman coding

The Huffman coding is a type of prefix code that uses character frequencies to assign a code to a variable. The most frequent character gets the smallest code, and the least frequent character gets the largest code [14]. The algorithm starts with the two minimum frequency variables and sums their frequency. Also, it is assigned an internal leaf code of 0 or 1 determined by its position. Then, the sum done becomes the frequency of a tree node. Next, the two following minimum frequencies are added, it could be the node we built before or different variables and, we continue building a tree until the tree only has one root. The algorithm complexity in the worst case is O(n log n).

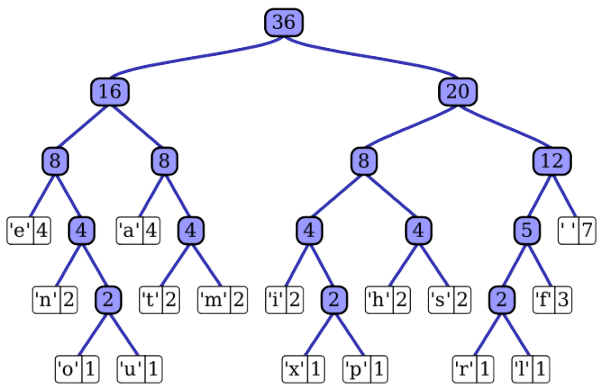

*Image 7. Huffman coding*

### 3.3.3 LZ77

Also called search buffer. It is a dictionary-based compression scheme. Codewords for substrings are pointers to the longest match for the substring found in the search buffer. When a matching appears, built a token formed by the steps to go back codeword for the substring, the length of the match, and the character of the next symbol. In software, the PNG file format is based on LZ77. In situations where a pattern repeats over a period larger than the search buffer size, the repetition cannot be taken advantage. The algorithm complexity in the worst case is O(n), for a text of n characters [15] .

*Image 8. Application LZ77 Algorithm*

### 3.3.4 LZW

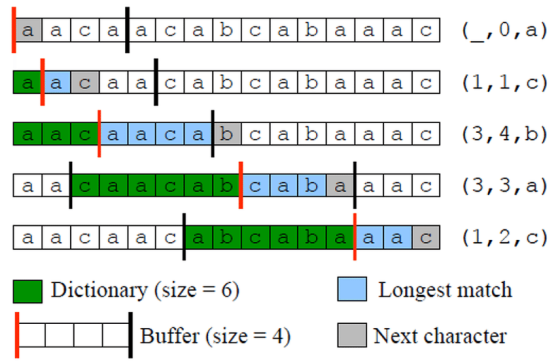

This algorithm is an improvement of the LZ78 algorithm. It is based on single-string character dictionary coders and is used to compress GIF images. It will maintain a running variable-length dictionary of symbols and trying to optimize for the longest match possible. As the message grows, however, the compression ratio tends asymptotically to the maximum [16]. It works by reading the first character string and check if it is in the dictionary. If not, it's added. Then check the second one and, also join the two first characters and add this joining as an extension of the dictionary to compress. The algorithm decompression works with the initial dictionary that contains a single-string character. The algorithm complexity is O(m,n), which m and n are the sizes of the bidimensional pixels array.

| Current Char | Next Char | CurrentChar + NextChar is in the dictionary ? | Output Index | [New Index] New String |
|---|---|---|---|---|
| "B" | "A" | No | 66 | [256]"BA" |
| "A" | "B" | No | 65 | [257]"AB" |
| "B" | "A" | Yes | - | - |
| "BA" | "A" | No | 256 | [258]"BAA" |
| "A" | "B" | Yes | - | - |
| "AB" | "A" | No | 257 | [259]"ABA" |
| "A" | "A" | No | 65 | [260]"AA" |
| "A" | "A" | Yes | - | - |
| "AA" | "A" | No | 260 | [261]"AAA" |
| "A" | EOF | - | 65 | - |

*Image 9. LZW Example*

## 4. ALGORITHM DESIGN AND IMPLEMENTATION

In what follows, we explain the data structures and the algorithms used in this work. The implementations of the data structures and algorithms are available at GitHub.

### 4.1 Data Structures

The data structure to make the image compression used was an importation from NumPy. It took the CSV file and transformed it into a NumPy array. Like it is shown in the following figures. Image 10 shows how the data file is brought to the code to be processed.

```
255,0,120,255,255,255,255,255,255,255
255,120,0,120,120,255,255,80,255,255
255,255,120,0,120,120,255,255,255,255
255,120,0,0,0,120,120,255,255,255
255,255,120,0,120,120,255,255,0,255
255,255,255,120,0,120,255,255,255,0
```

*Image 10. Data set in csv format*

Subsequently, the 11 image shows the transforming of the data after the NumPy function. It returns a NumPy array.

```
[[255,0,120,255,255,255,255,255,255,255],
 [255,120,0,120,120,255,255,80,255,255],
 [255,255,120,0,120,120,255,255,255,255],
 [255,120,0,0,0,120,120,255,255,255],
 [255,255,120,0,120,120,255,255,0,255],
 [255,255,255,120,0,120,255,255,255,0]]
```

*Image 11. Format of NumPy array*

### 4.2 Algorithms

In this work, we propose a compression algorithm which is a combination of a lossy image-compression algorithm and a lossless image-compression algorithm. We also explain how decompression for the proposed algorithm works.

For the compression of the images, we will first use a lossy compression algorithm to reduce the size of the image even if some information is lost; for this step, we use seam carving. This algorithm makes unions between the pixels that define a route in function of energy pixels compared with the nearest pixels to determine the less important which are the pixels with less energy and the be erased. Then we will proceed to use a lossy compression algorithm to further reduce the compression efficiency of the images and the time efficiency for the classification network through which the image will pass. The second algorithm to use is LZW which uses dictionary data structures to compress files, making comparison between strings and finding repeated patterns.

#### 4.2.1 Lossy image-compression algorithm

For the Lossy image-compression we decided to use seam carving because it is a really interesting algorithm. What the algorithm it creates an energy map from the original image. Then, it looks for seams with the least energy to remove them from the image. Usually, a big portion of these seams belong to the sky or dark spots. To apply the algorithm we firstly, transform the data from csv files to NumPy array with a NumPy function and we adjust the number of seams to erase. Then the algorithm returns the NumPy array compressed (with the seams erased) and we proceed to transform it to csv file and save to the computer to send it to lossless image compression algorithm.

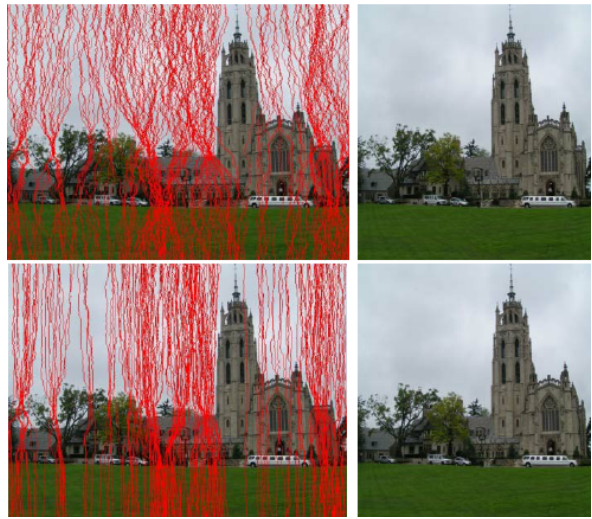

*Image 12. Seam carving (seams on the left)*

#### 4.2.2 Lossless image-compression algorithm

When using the LZW algorithm for image compression, we create a dictionary where we are going to have the image compressed. First, it creates a key for the first element of the array we are compressing (due to the nature of the algorithm, we must turn the 2D array into a flat list to then be processed). Then, it goes looking for equal sequences so if an arbitrary sequence with length x has appeared before, it storages full sequence. If an element has not appeared yet in the dictionary, it creates a new key. These two steps repeat until all the array has been iterated. Values for the dictionary are the index of found elements.

|    | Compressed Output | Dictionary | Buffer | Uncompressed Input |
|----|------|--------|---|-----------|
| a) |      |        | 0 | 100110101 |
| b) | 0    | 2(0,1) | 0 | 100110101 |
| c) | 01   | 3(1,0) | 1 | 00110101  |
| d) | 010  | 4(0,0) | 0 | 0110101   |
| e) | 010  |        | 0 | 110101    |
| f) | 0102 | 5(0,1,1)| 2 | 10101    |
| g) | 0102 |        | 1 | 0101      |
| h) | 01023| 6(1,0,1)| 3 | 101      |
| i) | 01023|        | 1 | 01        |
| j) | 01023|        | 3 | 1         |
| k) | 010236|       | 6 |           |

*Image 13. LZW Algorithm*

Decompression works taking indexes and adding values to the list according to the number value of each index. We go from the start of the dictionary to the end to convert the dictionary back into a flat list.

### 4.3 Complexity analysis of the algorithms

To calculate the complexity compression, we analyze the complexity in the first compression algorithm and then in the second one. The Seam carving complexity was O(N, M), where N was the number of rows and the M was the number of columns. Subsequently, the LZW algorithm presents an O(Z) complexity where Z is $N^2 \ast M^2$ which are the worst case if the algorithm does not find repeated patterns and must append to the dictionary each pattern. M*N are the rows and columns of the bidimensional array.

Now, we multiply, O(N,M) by O($N^2$, $M^2$) for the compression algorithm.

To decompress N are the number of the elements of the one-dimensional array that receives and, also it has an append.

| Algorithm | Time Complexity |
|---|---|
| Compression | O($N^3 \ast M^3$) |
| Decompression | O($N^2$) |

**Table 2:** Time Complexity of the image-compression and image-decompression algorithms. Where N are the rows and M are the columns of the bidimensional array.

| Algorithm | Memory Complexity |
|---|---|
| Compression | O($N^3 \ast M^3$) |
| Decompression | O($N^2$) |

**Table 3:** Memory Complexity of the image-compression and image-decompression algorithms. Where N are the rows and M are the columns of the bidimensional array.

### 4.4 Design criteria of the algorithm

We decided to use seam carving algorithm because of the freedom the user owns to take certain number of seams an eliminate them from the original image. It makes possible to adjust the ratio compression to any number we choose.

The LZW algorithm was chosen because it uses dictionaries as its data structure and the dictionary insertion and accessing process has a O(1) complexity, the above, makes the algorithm more efficient. However, the algorithm needs to storage data on simple python vectors where adding a value (append) goes for O(n) complexity, where n is the size of the vector. We take values from another vector with size m so it results on a final complexity of O(n*m). If we compare this data structure with others used in different algorithms, we notice it is a good option to use.

Also, both algorithms are common around the world, so the people who want to use our researching to apply in their projects can be modified easier than if we would use others algorithm more complex.

Our code structure was designed as follows. Firstly, we wrote the corresponding importations to implement some functions we used. Then, we implemented the seam carving algorithm after transforming with a NumPy function the CSV files to NumPy array to make it compatible. When the NumPy array comes out seam carving it is saved as a CSV file again with the same function. Now, the algorithm has passed through a lossy compression algorithm and will be implemented in the lossy compression algorithm. Afterward, the compressed CSV file is sent to the LZW algorithm, and it returns as a compressed file with lzw format and it can be decompressed and returned as a CSV file. We call all the functions in the main method applying our code implementation to all files or the number the user wants to read only changing the interval in the cycle we made. It works to read healthy and sick cattle files at the same time.

## 5. RESULTS

### 5.2 Execution times

In what follows we explain the relation of the average execution time and average file size of the images in the data set, in Table 6.

Compute execution time for each image in GitHub. Report average execution time Vs average file size.

|  | *Average execution time (s)* | *Average file size (MB)* |
|---|---|---|
| *Compression* | 13.75 s | 3.31 MB |
| *Decompression* | 7.75 s | 1.73 MB |

**Table 6:** Execution time of *Seam carving & LZW* algorithms for different images in the data set.

### 5.3 Memory consumption
We present memory consumption of the compression and decompression algorithms in Table 7.

|  | *Average memory consumption (MB)* | *Average file size (MB)* |
|---|---|---|
| Compression | 27 MB | 2.15 MB |
| Decompression | 19 MB | 1.27 MB |

**Table 7:** Average Memory consumption of all the images in the data set for both compression and decompression.

### 5.3 Compression ratio
We present the average compression ratio of the compression algorithm in Table 8.

|  | *Healthy Cattle* | *Sick Cattle* |
|---|---|---|
| Average compression ratio | 1.82:1 | 1.64:1 |

**Table 8:** Rounded Average Compression Ratio of all the images of Healthy Cattle and Sick Cattle.

### 6. DISCUSSION OF THE RESULTS
Applying both seam carving and LZW compression algorithms, we obtain the results we were looking for at the beginning of the investigation. Compressing cattle photos makes easier the process of sending and analyzing this data. The times we got for compression of all files including sick and healthy cattle accord to an affordable time. The more we compress the files the slower compression process becomes but the treatment after compression is better. We must respect some limits for the image to be recognizable and analyzed. With the ratio compression we got (1.82:1), the images are still totally useful for the objectives we had.

### ACKNOWLEDGEMENTS
This work was possible thanks to Universidad EAFIT, the government and their scholarships which allowed us to be here finishing this investigation started at the beginning of the semester.